\documentclass[twocolumn,showpacs,preprintnumbers,amsmath,amssymb]{revtex4}
\usepackage{graphicx}
\usepackage{dcolumn}
\usepackage{bm}
\usepackage{mathrsfs}
\usepackage{amsmath,amssymb,graphics}

\begin{document}

\title{Explicit field realizations of $W$ algebras}

\author{Shao-Wen Wei$^1$\footnote{E-mail: weishaow06@lzu.cn} ,
        Yu-Xiao Liu$^1$\footnote{Corresponding author. E-mail:
                             liuyx@lzu.edu.cn},
        Li-Jie Zhang$^2$\footnote{lijzhang@shu.edu.cn}
        and Ji-Rong Ren$^1$}
\affiliation{
    $^1$Institute of Theoretical Physics, Lanzhou University,
           Lanzhou 730000, P. R. China;\\
    $^2$Department of Physics, Shanghai University,
           Shanghai 200444, P. R. China.       }

\begin{abstract}
The fact that certain non-linear $W_{2,s}$ algebras can be
linearized by the inclusion of a spin-1 current can provide a
simple way to realize $W_{2,s}$ algebras from linear $W_{1,2,s}$
algebras. In this paper, we first construct the explicit field
realizations of linear $W_{1,2,s}$ algebras with double-scalar and
double-spinor, respectively. Then, after a change of basis, the
realizations of $W_{2,s}$ algebras are presented. The results show
that all these realizations are Romans-type realizations.

\end{abstract}


\maketitle

\section{Introduction}
\label{secIntroduction}

After the fundamental work of Zamolodchikov
\cite{Zamolodchikov1985} in the middle of the 1980's, $W$ algebras
have attracted much attention since they uncover some underlying
world-sheet symmetries of strings. Many $W$ algebras are known
(for review see \cite{BouwknegtPhysRept}) and much work has been
carried out on their classification
\cite{BoerCMP1994,DeckmynPRD1995,BoreschNPB1995,MadsenCMP1997}.
$W$ algebras have many applications and become the subject of
great interest in many branches of physics and mathematics, e.g.,
in $W$ gravity theories \cite{Pope1991,BergshoeffPLB1990},
critical and non-critical $W$ string theories
\cite{BergshoeffNPB1994,PopePLB1991},
Wess-Zumino-Novikov-Witten(WZNW) models
\cite{FeherPR1992,SevrinPLB1993,Frappatcmp1993}, quantum Hall
effect \cite{KarabaliNPB1994}, and especially in black holes
\cite{IsoPRD2007,BonoraJHEP2008}, where it was shown that the
Hawking radiation can be explained as the fluxes of chiral
currents forming a $W_\infty$ algebra.

As we know, $W$ algebras arise from Kac-Moody algebras, which are
related to classical Lie algebras. Various free field realizations
of $W$ algebras have been extensively studied
\cite{Ahnmpla1993,Frenkel1994,Howeplb1994,Boermpla1999,Wakimoto1998,Dotsenkoplb2001,Zhang0711.4624,IsaevaJMP2008}.
At quantum level, $W$ algebras are usually non-linear, which makes
it very difficult to give the field realizations of them. The
corresponding $W$ strings were first investigated in Ref.
\cite{GervaisNPB1987} and have been extensively developed since
then. Much research on the scalar realizations of $W_{2,s}$
strings has been done
\cite{RomansNPB1991,ThierryPLB1987,BaisNPB1988,SchransNPB1993,LuCQG1994,Lu2IJMPA1994,
Lu3CQG1994,BershadskyPLB1992,BergshoeffPLB1993}. Most of which are
based on the grading method, where the BRST charge of $W_{2,s}$
strings is written in the form of $Q_{B}=Q_{0}+Q_{1}$. This
provides an easy way to construct $W_{2,s}$ strings, while it
imposes more constrain conditions on the BRST charge. Under the
supposition that this grading form still holds true for spinor
field realizations, the corresponding works had been done
\cite{ZhaoPLB2000,DuanNPB2004,LiuJHEP2005,ZhangCTP2006}.

Furthermore, many investigations have been focused on
understanding the structure of $W$ algebras
\cite{KrivonosPLB1994,BellucciPLB1995,LuMPLA,Madsen1995}. It was
shown that linear Lie algebras with finite number of currents may
contain some non-linear $W$ algebras with an arbitrary central
charge as subalgebras. Especially for $W_{2,s}$ algebras, they can
be linearized by the inclusion of a spin-1 current at $s=3$ and 4
\cite{KrivonosPLB1994}. After performing a non-linear change of
basis, $W_{2,s}$ algebras can be recast into the form of linear
algebras. But for the spin-$s$ current $W_{0}$, one has
$W_{0}(z)W_{0}(\omega) \sim 0$, which indicates that $W_0$ is a
null current. It is exciting that this shines some light on the
realizations of the non-linear $W_{2,s}$ algebras. After
constructing the linear bases of $W_{1,2,s}$ algebras and making a
change of basis, we can obtain the realizations of the non-linear
$W_{2,s}$ algebras. In fact, the spin-$s$ current $W_{0}$ can be
set to zero, which will give a Romans-type realization of
$W_{2,s}$ algebras. However, in \cite{KrivonosPLB1994}, it was
shown that the null current $W_{0}$ does not need to be set to
zero and was first realized with parafermionic vertex operators.
It also can be found in \cite{LiuJHEP2005,BellucciPLB1995,LuMPLA}
that the null current $W_{0}$ was realized with the ghost-like
fields. In this paper, we will construct the linear bases of the
$W_{1,2,s}$ algebras with double-scalar and double-spinor,
respectively. Through a change of basis, we obtain some new
realizations of the non-linear $W_{2,s}$ algebras. All these
results show that there exists no non-Romans-type realization with
double-scalar or double-spinor only. However, we still expect that
there exist non-Romans-type realizations of $W$ algebras at some
special values of central charge.

The paper is organized as follows. In section \ref{Note on the
realization}, we give a brief review and analysis of the
realizations of the $W_{2,s}$ algebras and the $W_{2,s}$ strings.
Then in section \ref{Linearization}, we introduce the
linearization of the $W_{2,s}$ algebras. In sections \ref{Double
scalar fields} and \ref{Double spinor fields}, we construct the
bases of the linear $W_{1,2,s}$ algebras and obtain new
realizations of the $W_{2,s}$ algebras with double-scalar and
double-spinor, respectively. Finally, the paper ends with a brief
conclusion.

\section{Note on the realizations of the $W_{2,s}$ algebras and the $W_{2,s}$ strings}
\label{Note on the realization}

It is known that when extended to the quantum case, the $W_{2,s}$
algebras will become non-linear. The OPE of two currents with spin
$s$ and $s'$ produces terms with spin ($s+s'-2$) at leading order.
For example, there will be terms with spin-4 and spin-6 currents
in the OPEs of the $W_{2,3}$ algebra and the $W_{2,4}$ algebra,
respectively. However these terms with spin ($s+s'-2$) may be
interpreted as composite fields built from the products of the
fundamental currents with spin $s$ and $s'$. The $W_{2,3}$ algebra
is generated by the spin-2 energy-momentum tensor $T$ and spin-3
current $W$, which satisfy the OPEs \cite{Zamolodchikov1985}
\begin{eqnarray}
 T(z) T(\omega) & \sim & \frac{C/2}{(z-\omega)^{4}}+ \frac{2T}{(z-\omega)^{2}}
       +\frac{\partial T}{z-\omega}, \nonumber \\
 T(z) W(\omega) & \sim &  \frac{3 W}{(z-\omega)^{2}}
       +\frac{\partial W}{z-\omega}, \label{W23}\\
 W(z) W(\omega)
          & \sim &  \frac{C/3}{(z -w)^6}
                  +\frac{2T}{(z-\omega)^4}
                  + \frac{\partial T}{(z-\omega)^3} \nonumber\\
          &+& \frac{1}{(z-\omega)^2}\left(2 \Theta \Lambda+\frac{3}{10}\partial ^2T \right)\nonumber\\
          &+&\frac{1}{(z-\omega)}
                   \left(\Theta \partial \Lambda+\frac{1}{15}\partial ^3 T\right),\nonumber
\end{eqnarray}
where the coefficient $\Theta$ and composite field $\Lambda$ (spin
4) are given by
\begin{equation}
 \Theta= \frac{16}{22+5C},\;\;\;\;\;\; \Lambda = T^2
 - \frac{3}{10}
 \partial ^2 T.
\end{equation}
The constant $C$ is the central charge of the $W_{2,3}$ algebra.
It is easy to see that the denominator of $\Theta$ at
$C=-\frac{22}{5}$ will be zero and the $W_{2,3}$ algebra will
become singular. But one can rescale these currents such that the
corresponding OPEs are well defined, i.e., there have no divergent
coefficients in them (for the detailed discussion see
\cite{LuIPLB1995,wei2008}).

The $W_{2,4}$ algebra is given by \cite{KauschNPB1991}
\begin{eqnarray}
 && T(z) T(\omega) ~ \sim \frac{C/2}{(z-\omega)^{4}}+\frac{2T}{(z-\omega)^{2}}
       +\frac{\partial T}{z-\omega}, \nonumber \\
 && T(z) W(\omega)  \sim  \frac{4 W}{(z-\omega)^{2}}
       +\frac{\partial W}{z-\omega}, \nonumber\\
 && W(z)W(\omega) \label{W24} \\
 && \sim  \bigg\{\frac{2 T}{(z - w)^6}
   +\frac{\partial T}{(z-\omega)^5}
   + \frac{\frac{3}{10}\partial ^2 T +\sigma_1 U
      +\sigma_2 W}{(z-\omega)^4}  ~~~~~~  \nonumber\\
 &&+  \frac{1}{15}\frac{\partial ^3 T}{(z-\omega)^3}
   +\frac{1}{84}\frac{\partial ^4 T}{(z-\omega)^2}
   +\frac{1}{560}\frac{\partial ^5 T}{(z-\omega)}\bigg\}\nonumber\\
 &&+  \sigma_{1} \bigg\{\frac{1}{2} \frac{\partial U}{(z-\omega)^3}
   + \frac{5}{36} \frac{\partial ^2 U}{(z-\omega)^2}
   +\frac{1}{36} \frac{\partial ^3 U}{(z-\omega)} \bigg\}\nonumber\\
 &&+  \sigma_{2} \bigg\{\frac{1}{2}\frac{\partial W}{(z-\omega)^3}
   + \frac{5}{36} \frac{\partial ^2 W}{(z-\omega)^2}
   +\frac{1}{36} \frac{\partial ^3 W}{(z-\omega)} \bigg\}\nonumber\\
 &&+ \sigma_{3} \bigg\{ \frac{G}{(z-\omega)^2}
   +\frac{1}{2}\frac{\partial G}{(z-\omega)} \bigg\}\nonumber\\
 &&+ \sigma_{4}\bigg\{ \frac{A}{(z-\omega)^2}
   +\frac{1}{2}\frac{\partial A}{(z-\omega)}\bigg\}\nonumber\\
 &&+ \sigma_{5}\bigg\{ \frac{B}{(z-\omega)^2}
   +\frac{1}{2} \frac{\partial B}{(z-\omega)}\bigg\}
   +\frac{C/4}{(z-\omega)^8},\nonumber
\end{eqnarray}
where the composite fields $U$, $G$, $A$ and $B$ are defined by
\begin{equation}
\aligned
 U&=(TT)-\frac{3}{10} \partial ^2 T,\;\;\;\;\;\;\\
  G&=(\partial ^2 T
T)-\partial (\partial T T)+\frac{2}{9} \partial ^2
(TT)-\frac{1}{42}\partial ^4 T,\\
A & =(T U)-\frac{1}{6}\partial ^2 U, \;\;\;\;\;\; B=(T
W)-\frac{1}{6}\partial ^2 W ,
\endaligned
\end{equation}
with normal ordering of products of currents understood. The
coefficients $\sigma_{i}\;(i=1-5)$ are
\begin{equation}
\aligned
\sigma_{1} & =\frac{42}{5C+22},\;\;\;\;\;\;\\
\sigma_{2} & =\sqrt{\frac{54(C+24)(C^2 -172C
            +196)}{(5C+22)(7C+68)(2C-1)}},\\
\sigma_{3} & =\frac{3(19C-524)}{10(7C+68)(2C-1)},\;\;\;\;\;\;\\
\sigma_{4}
 & =\frac{24(72C+13)}{(5C+22)(7C+68)(2C-1)},\\
\sigma_{5} & = \frac{28}{3(C+24)} \sigma_{2}.\\
\endaligned
\end{equation}
It is worth to point out that, just as the $W_{2,3}$ algebra, the
$W_{2,4}$ algebra is singular at $C=-24$, $\frac{1}{2}$,
$-\frac{22}{5}$ and $-\frac{68}{7}$. After rescaling the spin-4
current $W$, it can be proved that only the case $C=-24$ satisfies
the Jacobi identity \cite{LuIPLB1995}.

In general, the BRST charge $Q_{B}$ for a $W_{2,s}$ string is
\cite{wei2008,BergshoeffPLB1995}
\begin{equation}
 Q_{B}=\oint dz \left[c(z)T(z)
             +\gamma (z) W(z)\right], \label{QB}
\end{equation}
where the currents $T$ and $W$ generate the corresponding
$W_{2,s}$ algebra, and the fermionic ghosts $(b,c)$ and
$(\beta,\gamma)$ are introduced for the currents $T$ and $W$,
respectively. It is easy to prove that the BRST charge given above
does satisfy the nilpotency condition:
\begin{equation}
 Q_{B}^{2}=\{Q_{B},Q_{B}\}=0.
\end{equation}

A realization for a $W_{2,s}$ algebra means giving an explicit
construction of the bases $T$ and $W$ from the basic fields, i.e.,
scalar fields, spinor fields, or ghost fields. Giving a
realization for a non-linear algebra is difficult and complex. For
simplicity, $Q_B$ can generally be expressed as the grading form
in many works:
\begin{eqnarray}
 Q_{B}&=&Q_{0}+Q_{1}, \label{gradingQB}\\
 Q_{0}&=&\oint dz c T, \\
 Q_{1}&=&\oint dz \gamma W, \label{Wconstruct}
\end{eqnarray}
where the currents $T$ and $W$ generate the $W_{2,s}$ algebras.
The detailed construction of the current $T$ can be found in
\cite{weiNPB2008}, where $T$ was constructed from scalar, spinor,
and ghost fields. Here the ghost fields $b$, $c$, $\beta$,
$\gamma$ are all fermionic and anticommuting. They satisfy the
OPEs
\begin{equation}
 b(z) c(\omega) \sim \frac{1}{z-\omega}, \quad \beta(z) \gamma(\omega) \sim
    \frac{1}{z-\omega},
\end{equation}
in other cases the OPEs vanish. The nilpotency condition of
$Q_{B}$ become
\begin{equation}
Q_{0}^{2}=Q_{1}^{2}=\{ Q_{0},Q_{1} \}=0. \label{nicon}
\end{equation}
Although it is easy to construct the $W_{2,s}$ strings in this
grading form, one may note that this gives more constraint
conditions on $Q_{B}$.

One also note that if we obtain a realization for a $W_{2,s}$
algebra, the BRST charge $Q_{B}$ of the corresponding $W_{2,s}$
string will be obtained by substituting the explicit forms of
currents $T$ and $W$ into (\ref{QB}).

\section{Linearization of the $W_{2,s}$ algebras from the $W_{1,2,s}$ algebras}
\label{Linearization}

It is shown that the $W_{2,s}$ algebras can be linearized as the
linear $W_{1,2,s}$ algebras generated by currents $J$, $T$ and $W$
with spin 1, 2 and $s$ respectively. The linear $W_{1,2,s}$
algebras for $s=3,4$ take the forms \cite{KrivonosPLB1994}
\begin{eqnarray}
 T_{0}(z) T_{0}(\omega) &\sim& \frac{C_{0}/2}{(z-\omega)^{4}}
   +\frac{2T}{(z-\omega)^{2}}+\frac{\partial T}{z-\omega}, \nonumber\\
 T_{0}(z) W_{0}(\omega) &\sim&  \frac{s W}{(z-\omega)^{2}}
   +\frac{\partial W}{z-\omega},\nonumber\\
 T_{0}(z)J_{0}(\omega) &\sim& \frac{C_{1}}{(z-\omega)^{3}}
   +\frac{J_{0}}{(z-\omega)^{2}}+\frac{\partial J_{0}}{z-\omega},\label{OPE}\\
 J_{0}(z)J_{0}(\omega) &\sim& -\frac{1}{(z-\omega)^{2}}, \nonumber  \\
 J_{0}(z)W_{0}(\omega) &\sim& \frac{\xi W_{0}}{z-\omega},\quad
 W_{0}(z)W_{0}(\omega) \sim 0. \nonumber
\end{eqnarray}
The coefficients $C_{0}$, $C_{1}$ and $\xi$ are given by
\begin{eqnarray} \label{c0 and c1}
\aligned
 &C_{0}=50+24t^{2}+\frac{24}{t^{2}},\quad
 C_{1}=-\sqrt{6}\big(t+\frac{1}{t}\big),\\
 &\xi=\sqrt{\frac{3}{2}}t ~~~~~~~~~~~~~~~~~~~~~~~~~~~~~~~~~~~~~(s=3),\\
 &C_{0}=86+30t^{2}+\frac{60}{t^{2}},\quad
 C_{1}=-3t-\frac{4}{t},\\
 &\xi=t ~~~~~~~~~~~~~~~~~~~~~~~~~~~~~~~~~~~~~~~~~~(s=4).
\endaligned
\end{eqnarray}
where $t$ is a non-zero constant. From these OPEs (\ref{OPE}), it
is clear that the current $W_{0}$ is a primary field with spin
$s$, while the current $J_{0}$ is not unless $C_{1}=0$. Here the
spin-$s$ current $W_{0}$ is null, and we will construct the most
general forms of it in next section. The results there show that
$W_{0}$ is zero. It also can be seen that every term on the right
hand side of the OPEs $T_{0}(z) W_{0}(\omega)$ and $J_{0}(z)W_{0}$
has $W_{0}$, so one can consistently set it to zero, though it
does not need to set to zero. In \cite{KrivonosPLB1994}, the null
current was first realized with parafermionic vertex operators,
and later was realized with the ghost-like fields
\cite{LiuJHEP2005,BellucciPLB1995,LuMPLA}.

The bases $T$ and $W$ of $W_{2,s}$ algebras were constructed by
the linear bases of the $W_{1,2,s}$ algebras in our previous paper
\cite{LiuJHEP2005}. For simplicity, we choose $t=-1$, $T$ and $W$
are given by
\begin{eqnarray}
 T &=&T_0 , \label{W2324T} \\
 W &=&W_{0}+\frac{7 i}{8} \partial^2 J_{0}- i \frac{\sqrt{6}}{2} \partial J_{0}J_{0}
     +\frac{i}{6} J_{0}^{3}\nonumber\\
     &&- i\frac{\sqrt{6}}{8} \partial T_{0}
     +\frac{i}{4} T_{0}J_{0}, \quad\quad\quad\quad ~~~~~~(s=3)  \label{W23X} \\
 W &=& W_0 + \frac{3a}{520} \partial^3 J_0 - \frac{3a}{260} \partial^2 J_0J_0
     - \frac{19a}{1560}(\partial J_0)^2\nonumber \\
     &&+ \frac{7a}{780} \partial J_0(J_0)^2
     -\frac{a}{1560}(J_0)^4
     -\frac{149}{390a} \partial ^2 T_0\nonumber \\
    && -\frac{59}{780a}(T_0)^2
     + \frac{a}{390} \partial T_0 J_0 + \frac{a}{260}T_0 \partial J_0\nonumber \\
    &&-\frac{a}{780} T_0(J_0)^2, ~~~~~~~~~~~~~~~~~~~~~~~~~~ (s=4) \label{W24X}
\end{eqnarray}
where $a=\sqrt{\frac{451}{2}}$. In this case, $T =T_{0}$ implies
that the central charges of the linear $W_{1,2,s}$ algebras and
the $W_{2,s}$ algebras are equal, i.e. $C=C_{0}$. One can also
shift it with an arbitrary constant $C_{eff}$, the center charge
of an effective energy-momentum tensor $T_{eff}$, and rewrite the
non-linear basis as $T=T_{0}+T_{eff}$. Here, we have shown that
$W_{1,2,s}$ algebras (\ref{OPE}) are linear and contain the
$W_{2,s}$ algebras as subalgebras. But one needs to keep in mind
that this linearization does not contain the case
$C=-\frac{22}{5}$ for $W_{2,3}$, and the cases $C=-24$,
$\frac{1}{2}$, $-\frac{22}{5}$ and $-\frac{68}{7}$ for $W_{2,4}$,
for which these algebras are singular.

\section{Double-scalar realizations for the linear $W_{1,2,s}$ algebras and the $W_{2,s}$ algebras }
\label{Double scalar fields}

In this section, we would like to construct the bases of the
linear $W_{1,2,s}$ algebras with double-scalar. Using the fact
that the $W_{2,s}$ algebras are contained in the linear
$W_{1,2,s}$ algebras as subalgebras, we can obtain new
realizations for the $W_{2,s}$ algebras by a change of basis.

\subsection{Realizations for the $W_{1,2,3}$ algebra and the $W_{2,3}$ algebra}

First of all, we notice the relation between $C_{0}$ and $C_{1}$
for $s=3$ shown in (\ref{c0 and c1}):
\begin{equation}
 C_{0}=2+4 C_{1}^{2}.
\end{equation}
The scalar field has spin 0 in conformal field theory, and the OPE
of it with itself is given by
\begin{equation}
 \phi (z) \phi(\omega) \sim \ln(z-\omega),
\end{equation}
or expressed as
\begin{equation}
 \partial \phi (z) \partial \phi(\omega) \sim
       -\frac{1}{(z-\omega)^{2}}.
\end{equation}
One needs to note that the field $\phi$ here is real. If $\varphi$
is a complex scalar field, it is easy to prove that the OPE will
be of the form
\begin{equation}
 \partial \varphi^{\dag}  \partial \varphi \sim
       -\frac{1}{(z-\omega)^{2}},
\end{equation}
in other cases the OPEs vanish.

Now we consider two real scalar fields $\phi_{1}$ and $\phi_{2}$.
The OPEs of them with each other are read as
\begin{equation}
 \partial \phi_{i} (z) \partial \phi_{j}(\omega) \sim
       -\frac{\delta_{ij}}{(z-\omega)^{2}}, \quad \quad  \quad (i,j=1,2).
\end{equation}
We would like to construct the explicit forms for the linear bases
of the $W_{1,2,3}$ algebra. The most general form of the basis
$T_{0}$ can be expressed as
\begin{equation}
 T_{0} = T_{eff}+g_{1}T_{\phi_{1}}+g_{2}T_{\phi_{2}}+ g_{3}T_{\phi_{1}\phi_{2}},
\end{equation}
where $T_{eff}$ is an effective energy-momentum tensor with
central charge $C_{eff}$. The introduction of $T_{eff}$ will
ensure the nontriviality of the solutions. $T_{\phi_{1}}$ and
$T_{\phi_{2}}$ are spin-2 energy-momentum tensors constructed from
fields $\phi_{1}$ and $\phi_{2}$, respectively, and
$T_{\phi_{1}\phi_{2}}$ is constructed from these two scalar
fields. The construction is
\begin{eqnarray}
 &&T_{\phi_{1}} =-\frac{1}{2}(\partial \phi_{1})^{2}- q_{1}\partial^{2} \phi_{1} ,  \\
 &&T_{\phi_{2}} =-\frac{1}{2}(\partial \phi_{2})^{2}- q_{2}\partial^{2} \phi_{2} ,  \\
 &&T_{\phi_{1}\phi_{2}} = \partial \phi_{1} \partial \phi_{2},
\end{eqnarray}
where $q_{1}$ and $q_{2}$ are the background charges of
$T_{\phi_{1}}$ and $T_{\phi_{2}}$, respectively. The other two
linear bases are given by
\begin{eqnarray}
 J_{0} &=& g_{4} \partial \phi_{1}+g_{5} \partial \phi_{2}, \\
 W_{0} &=&g_{6}  \partial T_{eff}+g_{7} T_{eff} \partial \phi_{1}
     +g_{8} T_{eff} \partial \phi_{2}+g_{9} \partial^{3} \phi_{1} ~~~~\nonumber\\
   &+&g_{10}(\partial\phi_{1})^{3}
     +g_{11} \partial^{2} \phi_{1}\partial\phi_{1}
     +g_{12} \partial^{3} \phi_{2}\nonumber\\
     &+&g_{13} (\partial\phi_{2})^{3}
     +g_{14} \partial^{2} \phi_{2}\partial \phi_{2}
     +g_{15} (\partial \phi_{1})^{2} \partial \phi_{2}\nonumber\\
     &+&g_{16} \partial \phi_{1}(\partial \phi_{2})^{2}
     +g_{17} \partial^{2} \phi_{1}\partial \phi_{2}
     +g_{18} \partial \phi_{1}\partial^{2} \phi_{2}.
\end{eqnarray}
Plugging these linear bases into the OPEs relations (\ref{OPE}),
we could obtain all the coefficients. One can see that the
constant $t$ appeared in (\ref{c0 and c1}) does not take zero,
which determines $\xi \neq 0$. This leads to a main result
\begin{equation}
g_{i}=0 \quad \quad \text{for} \;\;i=6-18,
\end{equation}
which means that the current $W_{0}$ is zero. After carefully
calculation, we obtain two solutions:

\begin{itemize}

{\item Solution 1}
\begin{equation}
\begin{array}{l} \label{solution1}
  g_{1}=g_{2}=1, \;\;\;
  g_{3}=0,\;\;\;
  g_{4}=g_{5}=\frac{\sqrt{2}}{2} h,\;\;\;\\
  C_{1}=2 \sqrt{2} h,\;\;\;
  C_{eff}=8,\;\;\;
  C_{0}=34,\;\;\;\\
  q_{1}=q_{2}=-1,
\end{array}\nonumber
\end{equation}

{\item Solution 2}
\begin{equation}
\begin{array}{l}
  g_{1}=g_{2}=-g_{3}=\frac{1}{2},\;\;\;
  g_{4}=g_{5}=\frac{\sqrt{2}}{2} h,\;\;\;\\
  C_{1}=2 \sqrt{2} h,\;\;\;
  C_{eff}=9,\;\;\;
  C_{0}=34,\;\;\;\\
  q_{1}=q_{2}=-2,
\end{array}\nonumber
\end{equation}
\end{itemize}
where $h$ satisfies $h^{2}=1$. The main difference between above
two solutions is whether the energy-momentum tensor
$T_{\phi_{1}\phi_{2}}$ vanishes. In Solution 1, the term
$T_{\phi_{1}\phi_{2}}$ does not appear. However, in Solution 2,
the contribution of the term $T_{\phi_{1}\phi_{2}}$ to central
charge is $\frac{1}{2}$.

Having found two realizations of the linear $W_{1,2,3}$ algebra,
we substitute the exact forms of the linear bases $T_{0}$ and
$J_{0}$ into (\ref{W23X}) and obtain two new realizations of the
$W_{2,3}$ algebra. The first realization is
\begin{eqnarray}
 T &=&T_{eff}+ (\partial \phi_{1})^{2}-\frac{1}{2}\partial^{2} \phi_{1}
          +(\partial \phi_{2})^{2}-\frac{1}{2}\partial^{2} \phi_{2},\\
 W &=& \frac{\sqrt{2}i}{48}
      \bigg( 6h\; T_{eff}\partial \phi_{1}+6h \; T_{eff}\partial \phi_{2}-h \; (\partial
            \phi_{1})^{3} \nonumber\\
      &&    +3h\; (\partial \phi_{1})^{2}\partial \phi_{2}
            +9h\; \partial \phi_{1}(\partial \phi_{2})^{2}\nonumber\\
      &&    +(6h-12\sqrt{3})\; \partial \phi_{1}
              \partial^{2} \phi_{2}-h\;(\partial \phi_{2})^{3} \nonumber\\
      &&    +(6h-6\sqrt{3})\;\partial^{2} \phi_{1} \partial \phi_{1}
            +(6h-12\sqrt{3}) \partial^{2}\phi_{1}\partial \phi_{2} ~~~~\nonumber\\
      &&    +(6h-6\sqrt{3})\; \partial^{2}\phi_{2}\partial\phi_{2}
            -6\sqrt{3}\;\partial T_{eff}\nonumber\\
      &&    +(24h-6\sqrt{3})\partial^{3}\phi_{1}
            +(6h-6\sqrt{3})\partial^{3}\phi_{2}\bigg),
\end{eqnarray}
and the second one reads
\begin{eqnarray}
 T &=&T_{eff}+ (\partial \phi_{1})^{2}-\frac{1}{2}\partial^{2} \phi_{1}
          +(\partial \phi_{2})^{2} \nonumber\\
      &&  -\frac{1}{2}\partial^{2} \phi_{2}
          -\frac{1}{2}\partial \phi_{1} \partial \phi_{2}, \\
 W &=& \frac{\sqrt{2}i}{96}
       \bigg( 12h\; T_{eff}\partial \phi_{1}
            +12h \; T_{eff}\partial \phi_{2}\nonumber\\
      &&    +3h\; \partial \phi_{1}(\partial \phi_{2})^{2}
            +(12\sqrt{3}+12h)\;\partial^{2} \phi_{1} \partial \phi_{1} \nonumber\\
      &&    +(12h-18\sqrt{3})\; \partial \phi_{1} \partial^{2} \phi_{2}
            +h\;(\partial \phi_{2})^{3}\nonumber\\
      &&    -2h \; (\partial \phi_{1})^{3}
            +(12h-18\sqrt{3}) \partial^{2}\phi_{1}\partial \phi_{2}\nonumber\\
      &&    +(12h-18\sqrt{3})\; \partial^{2}\phi_{2}\partial\phi_{2}
            -12\sqrt{3}\;\partial T_{eff}\nonumber\\
      &&    +(51h-12\sqrt{3})\;\partial^{3}\phi_{1}
      +(48h-12\sqrt{3})\;\partial^{3}\phi_{2}\bigg),~~~~~
\end{eqnarray}
where $h$ satisfies $h^{2}=1$. Note that, although
$T_{\phi_{1}\phi_{2}}$ is absent in the first realization, the
energy-momentum tensor $T$ in both realizations has central charge
$C=34$. If plugging these realizations into (\ref{QB}), one will get
the BRST charges for the $W_{2,3}$ string.

\subsection{Realizations for the $W_{1,2,4}$ algebra and the $W_{2,4}$ algebra}

For the linear $W_{1,2,4}$ algebra, the relation between $C_{0}$
and $C_{1}$ is
\begin{equation}
 C_{0}=1+\frac{1}{24}(85 C_{1}^{2}-5C_{1}h\sqrt{-48+C_{1}^{2}}).
\end{equation}
Next, we would like to construct the explicit forms of the linear
bases of the $W_{1,2,4}$ algebra. The most general forms of bases
$T_{0}$ and $J_{0}$ are
\begin{eqnarray}
 &&T_{0} =f_{1}T_{eff}+f_{2}T_{\phi_{1}}+f_{3}T_{\phi_{2}}+ f_{4}T_{\phi_{1}\phi_{2}}, \\
 &&J_{0} =f_{5} \partial \phi_{1}+f_{6} \partial \phi_{2},
\end{eqnarray}
where the energy-momentum tensors $T_{\phi_{1}}$, $T_{\phi_{2}}$
and $T_{\phi_{1}\phi_{2}}$ are given by
\begin{eqnarray}
 &&T_{\phi_{1}} =-\frac{1}{2}(\partial \phi_{1})^{2}- q_{3}\partial^{2} \phi_{1} ,  \\
 &&T_{\phi_{2}} =-\frac{1}{2}(\partial \phi_{2})^{2}- q_{4}\partial^{2} \phi_{2} ,  \\
 &&T_{\phi_{1}\phi_{2}} = \partial \phi_{1} \partial \phi_{2}.
\end{eqnarray}
For the linear basis $W_{0}$ with spin 4, the calculation shows
that $W_{0}\sim 0$. Under this case, the current $W$ of the
$W_{2,4}$ algebra is constructed from the linear bases $T_{0}$ and
$J_{0}$ only. Plugging these linear bases into the OPEs
(\ref{OPE}), we obtain two solutions, where the energy-momentum
tensor $T_{eff}$ vanishes in both cases. These solutions are
listed as follow:

\begin{itemize}

{\item Solution 1}
\begin{equation}
\begin{array}{l} \label{solution1}
  f_{1}=0,\;\;\;
  f_{2}=f_{3}=1,\;\;\;
  f_{4}=0,\;\;\;
  f_{5}=f_{6}=\frac{\sqrt{2}}{2} h,\;\;\;\\
  C_{1}=i\sqrt{2} h ,\;\;\;
  C_{0}=-4,\;\;\;
  q_{3}=q_{4}=-i\frac{h}{2},
\end{array}\nonumber
\end{equation}

{\item Solution 2}
\begin{equation}
\begin{array}{l}
  f_{1}=0,\;\;\;
  f_{2}=f_{3}=\frac{1}{2},\;\;\;
  f_{4}=-\frac{1}{2},\;\;\;
  f_{5}=f_{6}=\frac{\sqrt{2}}{2} h,\;\;\;\\
  C_{1}=\frac{5\sqrt{3}i}{3} h,\;\;\;
  C_{0}=-24,\;\;\;
  q_{3}=q_{4}=-\frac{5i}{6}h,
\end{array}\nonumber
\end{equation}
\end{itemize}
where $h^{2}=1$. It is clear that $f_{1}=0$ in both solutions and
this means the vanishing of the energy-momentum tensor $T_{eff}$.
Therefore, $C_{eff}=0$. The coefficient $f_{4}=0$ in Solution 1
implies that $T_{\phi_{1}\phi_{2}}$ vanishes, however, it does not
vanish in Solution 2. One may also note that the central charge
$C_{0}$ in both solutions is negative, which is different from the
case of the $W_{1,2,3}$ algebra. The background charges $q_{3}$
and $q_{4}$ of $T_{\phi_{1}}$ and $T_{\phi_{2}}$ are all imaginary
number.

After constructing the explicit forms of the linear bases $T_{0}$
and $J_{0}$, we would like to substitute them into (\ref{W24X})
and obtain new realizations for the $W_{2,4}$ algebra. The
realization constructed from Solution 1 is
\begin{eqnarray}
 T &=&-\frac{1}{2} (\partial \phi_{1})^{2}-\frac{1}{2} (\partial
        \phi_{2})^{2}-\frac{i h}{2}\partial^{2} \phi_{1}-\frac{i h}{2}\partial^{2} \phi_{2},\\
 W &=& b_{1} (\partial \phi_{1})^{4}+b_{2} (\partial\phi_{1})^{2}(\partial\phi_{2})^{2}
        +b_{3} (\partial\phi_{1})^{2} \partial^{2}\phi_{2}\nonumber \\
      &&+b_{4} \partial\phi_{1}\partial^{2}\phi_{2}\partial\phi_{2}
        +b_{5} \partial\phi_{1} \partial^{3}\phi_{2}
        +b_{6} (\partial \phi_{2})^{4}\nonumber \\
      &&+b_{7}\partial^{2}\phi_{1}(\partial \phi_{1})^{2}
        +b_{8} \partial^{2}\phi_{1}\partial \phi_{1}\partial \phi_{2}
        +b_{9} \partial^{2}\phi_{1}(\partial \phi_{2})^{2} \nonumber \\
      &&+b_{10} (\partial^{2}\phi_{1})^{2}
        +b_{11} \partial^{2}\phi_{1} \partial^{2}\phi_{2}
        +b_{12} \partial^{2}\phi_{2}(\partial\phi_{2})^{2}\nonumber \\
      &&+b_{13} (\partial^{2}\phi_{2})^{2}
        +b_{14} \partial^{3}\phi_{1} \partial \phi_{1}
        +b_{15} \partial^{3}\phi_{1} \partial \phi_{2} \nonumber \\
      &&+b_{16} \partial^{3}\phi_{2} \partial \phi_{2}
        +b_{17} \partial^{4}\phi_{1}
        +b_{18} \partial^{4}\phi_{2}.
\end{eqnarray}
These coefficients are
\begin{eqnarray}
 b_{1}&=&-\frac{59}{3120a}+\frac{a}{6420},\quad
 b_{2}=-\frac{59}{1560a}-\frac{a}{3120}, \nonumber \\
 b_{3}&=&\frac{a}{390\sqrt{2}}+\frac{59 i h}{1560a}-\frac{i a h}{3120},
 b_{4}=\frac{a}{156\sqrt{2}}-\frac{i a h}{1560},\nonumber\\
 b_{5}&=&-\frac{3a}{520}+\frac{i a h}{780\sqrt{2}},\quad
 b_{6}=-\frac{59}{3120a}+\frac{a}{6240},\nonumber \\
 b_{7}&=&\frac{59 i h}{1560 a}-\frac{i a h}{3120},\quad
 b_{8}=\frac{a}{156\sqrt{2}}-\frac{i a h}{1560},\nonumber \\
 b_{9}&=&\frac{a}{390\sqrt{2}}+\frac{59 i h}{1560a}-\frac{i a h}{3120},\nonumber \\
 b_{10}&=&\frac{149}{390a}-\frac{19a}{3120}+\frac{i a h}{520\sqrt{2}}+\frac{59}{3120a},\nonumber \\
 b_{11}&=&-\frac{19a}{1560}+\frac{i a h}{260\sqrt{2}}+\frac{59}{1560a},\nonumber \\
 b_{12}&=&\frac{59 i h}{1560a}-\frac{i a h}{3120},\nonumber \\
 b_{13}&=&\frac{149}{390a}-\frac{19a}{3120}+\frac{i a h}{520\sqrt{2}}+\frac{59}{3120a},
       \nonumber \\
 b_{14}&=&\frac{149}{390a}-\frac{3a}{520}+\frac{i a h}{780 \sqrt{2}},\quad
 b_{15}=-\frac{3a}{520}+\frac{i a h}{780\sqrt{2}},\nonumber \\
 b_{16}&=&\frac{149}{390a}-\frac{3a}{520}+\frac{i a h}{780\sqrt{2}},\quad
 b_{17}=\frac{3a}{520\sqrt{2}}-\frac{149 i h}{780a},\nonumber \\
 b_{18}&=&\frac{3a}{520\sqrt{2}}-\frac{149 i h}{780a}, \nonumber
\end{eqnarray}
where $a=\sqrt{\frac{451}{2}}$.

Solution 2 gives a realization of the linear $W_{1,2,4}$ algebra
with central charge $C_{0}=-24$, which is singular and could not
be used to construct the $W_{2,4}$ algebra. But it is indeed a
realization of the $W_{1,2,4}$ algebra.

\section{Double-spinor realizations for the linear $W_{1,2,s}$ algebras and the $W_{2,s}$ algebras }
\label{Double spinor fields}

In this section, we would like to construct the bases of the
linear $W_{1,2,s}$ algebras with double-spinor. The new
realizations for the $W_{2,s}$ algebras can be obtained after a
change of bases.

\subsection{Realizations for the $W_{1,2,3}$ algebra and the $W_{2,3}$ algebra}

A spinor field has spin $\frac{1}{2}$ in conformal field theory,
and the OPE $\psi (z) \psi(\omega)$ is given by
\begin{equation}
 \psi (z) \psi(\omega) \sim -\frac{1}{z-\omega}.
\end{equation}
We denote two real spinor fields as $\psi_{1}$ and $\psi_{2}$, and
they satisfy
\begin{equation}
 \psi_{i} (z) \psi_{j}(\omega) \sim
       -\frac{\delta_{ij}}{z-\omega}, \quad \quad  \quad (i,j=1,2).
\end{equation}
Next, we would like to construct the explicit forms of the linear
bases for the $W_{1,2,3}$ algebra. The most general forms of
$T_{0}$, $J_{0}$ and $W_{0}$ can be expressed as
\begin{eqnarray}
 T_{0} &=& T_{eff}+h_{1} T_{\psi_{1}}+h_{2} T_{\psi_{2}}
           + h_{3}T_{\psi_{1}\psi_{2}},\nonumber\\
 J_{0} &=& h_{4} \psi_{1} \psi_{2}, \\
 W_{0} &=& h_{5} \partial^{2}\psi_{1} \psi_{1}+ h_{6} \partial^{2}\psi_{2} \psi_{2}
           +h_{7} \partial^{2}\psi_{1} \psi_{2}\nonumber\\
         &&+h_{8} \psi_{1} \partial^{2} \psi_{2}
           +h_{9} \partial\psi_{1} \partial\psi_{2} \nonumber\\
         &&+h_{10} \partial T_{eff}
           +h_{11} T_{eff} \psi_{1} \psi_{2}. \nonumber
\end{eqnarray}
The energy-momentum tensors $T_{\psi_{1}}$ and $T_{\psi_{2}}$ with
spin 2 are constructed from $\psi_{1}$ and $\psi_{2}$
respectively, and $T_{\psi_{1}\psi_{2}}$ is constructed from these
two spinor fields. They are constructed as
\begin{eqnarray}
 &&T_{\psi_{1}} = \partial\psi_{1}\psi_{1}, \label{TT1}\\
 &&T_{\psi_{2}} = \partial\psi_{2}\psi_{2}, \\
 &&T_{\psi_{1}\psi_{2}} = \partial\psi_{1}\psi_{2}+ \psi_{1}
 \partial\psi_{2}. \label{TT2}
\end{eqnarray}
Plugging these linear bases into the OPE relations (\ref{OPE}), we
obtain the result:
\begin{equation}
\begin{array}{l}
  h_{1}=h_{2}=-\frac{1}{2}, \;\;\;
  h_{3}=0 ,\;\;\;
  h_{4}=1,\;\;\; \\
  h_{i}=0\;\;(i=5-11) ,\;\;\;
  C_{1}=0,\;\;\;
  C_{0}=2,\;\;\;
  C_{eff}=1.
\end{array} \nonumber
\end{equation}
In this solution, it can be seen that $T_{\psi_{1}\psi_{2}}$ and
$W_{0}$ vanish. The energy-momentum tensor $T_{eff}$ contributes
central charge 1.

After constructing the explicit forms of the linear bases $T_{0}$,
 $J_{0}$ and $W_0$, we substitute them into (\ref{W23X}) and obtain a
realization of the $W_{2,3}$ algebra as follows:
\begin{eqnarray}
 T &=& T_{eff}-\frac{1}{2} \partial \psi_{1}\psi_{1}-\frac{1}{2} \partial \psi_{2}\psi_{2},\nonumber\\
 W &=& \frac{i}{4} T_{eff}\psi_{1}\psi_{2}
       +\frac{7 i}{8}\psi_{1}\partial^{2}\psi_{2}
       +\frac{\sqrt{6}i}{16} \partial^{2}\psi_{1}\psi_{1}\nonumber\\
     &&+\frac{7i}{8}\partial^{2}\psi_{1}\psi_{2}
       +\frac{\sqrt{6}i}{16} \partial^{2}\psi_{2}\psi_{2}
       -\frac{\sqrt{6}i}{8}\partial T_{eff}.  \nonumber
\end{eqnarray}
It is worth remarking that the currents $T$ and $W$ above generate
$W_{2,3}$ algebra with central charge $C=2$.

\subsection{Realizations for the $W_{1,2,4}$ algebra and the $W_{2,4}$ algebra}

For the linear $W_{1,2,4}$ algebra, the bases take the following
form
\begin{eqnarray}
 T_{0} &=& T_{eff}+k_{1} T_{\psi_{1}}+k_{2} T_{\psi_{2}}
           + k_{3}T_{\psi_{1}\psi_{2}},\nonumber\\
 J_{0} &=& k_{4} \psi_{1} \psi_{2}, \\
 W_{0} &=& 0, \nonumber
\end{eqnarray}
where $T_{\psi_{1}}$, $T_{\psi_{2}}$ and $T_{\psi_{1}\psi_{2}}$
are given by (\ref{TT1})-(\ref{TT2}). This case give a precise
Romans realization of the $W_{1,2,4}$ algebra, where the basis
$W_{0}$ is set to zero.

Plugging these linear bases into the OPE relations (\ref{OPE}), we
obtain two solutions:

\begin{itemize}

{\item Solution 1}
\begin{equation}
\begin{array}{l}
  k_{1}=k_{2}=-\frac{1}{2}, \;\;\;
  k_{3}=-\frac{C_{1}}{2},\;\;\;
  k_{4}=-1,\\
  C_{eff}=\frac{1}{24}(13C_{1}^{2}+5C_{1}h\sqrt{C_{1}^{2}-48}),\\
  C_{0}=1+\frac{1}{24}(85C_{1}^{2}+5C_{1}h\sqrt{C_{1}^{2}-48}).
\end{array}
\end{equation}

{\item Solution 2}
\begin{equation}
\begin{array}{l}
  k_{1}=k_{2}=-\frac{1}{2}, \;\;\;
  k_{3}=\frac{C_{1}}{2},\;\;\;
  k_{4}=1,\\
  C_{eff}=\frac{1}{24}(13C_{1}^{2}+5C_{1}h\sqrt{C_{1}^{2}-48}),\\
  C_{0}=1+\frac{1}{24}(85C_{1}^{2}+5C_{1}h\sqrt{C_{1}^{2}-48}).
\end{array}
\end{equation}
\end{itemize}
The central charges $C_{0}$ of both solutions depend on $C_{1}$,
and this gives realizations of the linear $W_{1,2,4}$ algebra at
arbitrary central charge. Then two new realizations of the
$W_{2,4}$ algebra from these solutions can be obtained
immediately. The first one is given by
\begin{eqnarray}
 T &=& T_{eff}-\frac{1}{2} \partial \psi_{1}\psi_{1}-\frac{1}{2} \partial \psi_{2}\psi_{2}  \nonumber\\
      &&-\frac{C_{1}}{2} \partial \psi_{1}\psi_{2}-\frac{C_{1}}{2} \psi_{1}\partial\psi_{2},\\
 W &=&-\frac{1}{1560a}\bigg(
        -2(3a^{2}-59C_{1})T_{eff}\psi_{1}\partial\psi_{2}\nonumber\\
      &&+118T_{eff}\partial\psi_{1}\psi_{1}
        -2(3a^{2}-59 C_{1})T_{eff}\partial\psi_{1}\psi_{2}\nonumber\\
      &&+118C_{1}T_{eff}\partial\psi_{2}\psi_{2}
        -9(a^{2}-32 C_{1})\psi_{1}\partial^{3}\psi_{1} \nonumber\\
      &&-4a^{2}\partial T_{eff} \psi_{1}\psi_{2}
        +298 \partial^{2}\psi_{2}\partial\psi_{2}\nonumber\\
      &&-(59+2a^{2}-2a^{2}C_{1}^{2}+59C_{1}^{2})\partial\psi_{1}\psi_{1}\partial\psi_{2}\psi_{2}\nonumber\\
      &&-(27a^{2}-894C_{1})\partial\psi_{1}\partial^{2}\psi_{2} \nonumber\\
      &&+298 \partial^{2}\psi_{1}\partial\psi_{1}
        -(27a^{2}-894C_{1}) \partial^{2}\psi_{1}\partial\psi_{2}\nonumber\\
      &&+298 \partial^{3}\psi_{1} \psi_{1}
        +(9a^{2}-298C_{1}) \partial^{3}\psi_{1} \psi_{2}\nonumber\\
      &&+298 \partial^{3}\psi_{2} \psi_{2}
        -596 \partial^{2} T_{eff}-118 T_{eff}^{2}\bigg).
\end{eqnarray}
The second is
\begin{eqnarray}
 T &=& T_{eff}-\frac{1}{2} \partial \psi_{1}\psi_{1}-\frac{1}{2} \partial \psi_{2}\psi_{2}\nonumber\\
      &&-\frac{C_{1}}{2} \partial \psi_{1}\psi_{2}-\frac{C_{1}}{2} \psi_{1}\partial\psi_{2},
\end{eqnarray}
\begin{eqnarray}
 W &=&-\frac{1}{1560a}\bigg(
        +2(3a^{2}-59C_{1})T_{eff}\psi_{1}\partial\psi_{2}\nonumber\\
      &&+118T_{eff}\partial\psi_{1}\psi_{1}
        +2(3a^{2}-59 C_{1})T_{eff}\partial\psi_{1}\psi_{2}\nonumber\\
      && +118C_{1}T_{eff}\partial\psi_{2}\psi_{2}
        +9(a^{2}-32 C_{1})\psi_{1}\partial^{3}\psi_{1} \nonumber\\
      &&-4a^{2}\partial T_{eff} \psi_{1}\psi_{2}
        +298 \partial^{2}\psi_{2}\partial\psi_{2} \nonumber\\
      &&-(59+2a^{2}-2a^{2}C_{1}^{2}+59C_{1}^{2})\partial\psi_{1}\psi_{1}\partial\psi_{2}\psi_{2} \nonumber\\
      &&+(27a^{2}-894C_{1})\partial\psi_{1}\partial^{2}\psi_{2} \nonumber\\
      &&+298 \partial^{2}\psi_{1}\partial\psi_{1}
        -(27a^{2}-894C_{1}) \partial^{2}\psi_{1}\partial\psi_{2} \nonumber\\
      &&+298 \partial^{3}\psi_{1} \psi_{1}
        -(9a^{2}-298C_{1}) \partial^{3}\psi_{1} \psi_{2} \nonumber\\
      &&+298 \partial^{3}\psi_{2} \psi_{2}
        -596 \partial^{2} T_{eff}-118 T_{eff}^{2}\bigg).
\end{eqnarray}
Different from the cases of scalar realizations, the results here
give two spinor realizations of the $W_{2,4}$ algebra for an
arbitrary central charge.

\section{Conclusion}
\label{secConclusion}

In this paper, we obtained the explicit field realizations of the
linear $W_{1,2,s}$ algebras and the non-linear $W_{2,s}$ algebras
with double-scalar and double-spinor, respectively. Owing to the
intrinsic nonlinearity of $W_{2,s}$ algebras, it is hard to
construct their field realizations. However, it is proved that the
non-linear $W_{2,s}$ algebras are contained in the linear
$W_{1,2,s}$ algebras with three currents $J_{0}$, $T_{0}$ and
$W_{0}$ as a subalgebra. With this fact, we first constructed the
linear bases of the $W_{1,2,s}$ algebras. Then making a change of
basis, we obtained several explicit field realizations of the
non-linear $W_{2,s}$ algebras. All these results imply a symmetry
under $\phi_{1} \leftrightarrow \phi_{2}$ or $\psi_{1}
\leftrightarrow \psi_{2}$. This method overcomes the difficulty of
realizations for a non-linear algebra.

The spin-$s$ current $W_{0}$ of $W_{1,2,s}$ algebras was
considered to be a null current and can be set to zero, then the
realizations of $W_{2,s}$ algebras obtained from linear
$W_{1,2,s}$ algebras are called Romans-type realizations. In fact,
it is not necessary to set the current $W_{0}$ to zero. In our
constructions, we first listed the most general forms of linear
bases $J_{0}$, $T_{0}$ and $W_{0}$ with correct spin. Plugging
these forms into the OPEs (\ref{OPE}), we found that all the
coefficients of $W_{0}$ vanished for the non-zero constant $\xi$.
These results suggest that there exists no non-Romans-type
realization of $W_{2,s}$ algebra if we use double-scalar or
double-spinor only. However, we expect that there exist
non-Romans-type realizations of $W_{2,s}$ algebras at some value
of central charge and more details would be investigated in our
future work.

We can also see that all these realizations satisfy $C=C_{0}$,
i.e., the central charge of the $W_{2,s}$ algebras are equal to
the central charge of the $W_{1,2,s}$ algebras, which is caused by
the assumption $T=T_{0}$. The central charge $C$ takes some
special values for the double-scalar realizations and the
double-spinor realizations of the $W_{2,3}$ algebras, while it
depends on the value of $C_{1}$for the double-spinor realizations
of the $W_{2,4}$ algebra. We also showed that there is no such
realization for the $W_{2,3}$ algebra at central charge
$C=-\frac{22}{5}$ and for the $W_{2,4}$ algebra at $C=-24$,
$\frac{1}{2}$, $-\frac{22}{5}$, and $-\frac{68}{7}$, since the
$W_{2,s}$ algebras are singular at these values of central charge.

\section*{Acknowledgements}

This work was supported by Program for New Century Excellent
Talents in University, the National Natural Science Foundation of
China (No. 10705013), the Doctoral Program Foundation of
Institutions of Higher Education of China (No. 20070730055), the
Key Project of Chinese Ministry of Education (No. 109153) and the
Fundamental Research Fund for Physics and Mathematics of Lanzhou
University (No. Lzu07002). L.J. Z acknowledges financial support
from Innovation Foundation of Shanghai University.

\end{document}